\documentclass[reprint,amsmath,amssymb,aps,floatfix]{revtex4-2}

\usepackage{graphicx}
\usepackage{dcolumn}
\usepackage{bm}

\usepackage{hyperref}
\hypersetup{breaklinks=true}
\usepackage[mathlines]{lineno}
\usepackage{url}
\usepackage{verbatim}
\usepackage{amsmath,amssymb}
\usepackage[normalem]{ulem}
\usepackage[usenames,dvipsnames]{color}

\begin{document}

\title{Comprehensive Study of Radon Progeny Attachment to Surfaces}

\author{D. Chernyak}
\author{J. Howell}
\author{D. Majumdar}
\author{N. Mukherjee} 
\thanks{Now at: Department of Physics and Astronomy, University of Tennessee, Knoxville, TN 37996, USA}
\author{O. Nusair}
\author{A. Piepke}

\affiliation{%
Department of Physics and Astronomy, University of Alabama, Tuscaloosa, AL 35487, USA\\
}%

\date{May 10, 2023}

\begin{abstract}
Low energy, low rate experiments, such as searches for neutrinoless double beta decay and dark matter, require unprecedentedly low levels of background in order to deliver their full science potential. $^{210}$Po driven, neutron induced background, caused by nuclear $(\alpha, n)$-reactions on low-Z materials, direct background contributions of the $^{210}$Po $\alpha$-radiation and desorption of the $^{210}$Pb progeny $^{210}$Bi from surfaces into the detector medium are of particular of concern. These backgrounds depend on details of the components' exposure to radon-loaded lab air and, thus, their handling history. The attachment rates of airborne radon progeny to surfaces, needed for the estimation of these background rates, are poorly understood. This article reports the results of a campaign comprising of more than 1200 attachment measurements, performed for 9 different materials. Correlations of the attachment with environmental parameters such as air exchange rate, electrical surface potential, temperature, atmospheric pressure, and relative humidity have been studied and found to be significant only in case of the first two. Attachment modelling, using the Jacobi model, is compared to data.
\end{abstract}

\keywords{RADIOACTIVITY}

\maketitle

\section{Introduction }
Next generation experiments, planning to explore neutrinoless double beta decay~\cite{cupid_pre-cdr_2019,legend_2021,nEXO:2022nam}, dark matter~\cite{Darwin_2016,darkside_2018} and solar neutrinos~\cite{juno_2016} rely on achieving unprecedented background rates to reach their scientific goals. 
$(\alpha, n)$-reactions, creating energetic neutrons, are an important source of background for them. The fast neutrons generated in these reactions are penetrating deep even into large detectors. This type of nuclear reaction may be driven by the $\alpha$-decay of $^{210}$Po, sustained by its long-lived parent $^{210}$Pb ($\rm T_{1/2}=22.20\; yr$). $^{210}$Pb surface radioactivity is a result of exposure to radon-containing lab air. Depending on the application, the $^{210}$Po $\alpha$ radiation may contribute directly to the detector background. Reference~\cite{KamLAND_7Be_2014} reported the desorption of the $^{210}$Pb progeny $^{210}$Bi as an important background component.

The radon progeny powered background, thus, depends on the handling history of experiment components and constitutes a memory effect. The relevant decay sequence is shown below, with the decay type and mean lifetimes given.
\begin{eqnarray*}
^{222}{\rm Rn}\; & 
\substack{5.52\; d \\ \xrightarrow{\makebox[0.6cm]{}} \\ \alpha} & \; 
^{218}{\rm Po}\; \substack{4.47\; m\\ \xrightarrow{\makebox[0.6cm]{}}\\ \alpha} \; ^{214}{\rm Pb}\; 
\substack{39.0\; m\\ \xrightarrow{\makebox[0.6cm]{}} \\ \beta} \; ^{214}{\rm Bi}\; 
\substack{28.7\; m\\ \xrightarrow{\makebox[0.6cm]{}}\\ \beta} \; ^{214}{\rm Po} \nonumber \\
  &  & \nonumber \\
^{214}{\rm Po}\; &
\substack{236\; \mu s\\ \xrightarrow{\makebox[0.6cm]{}}\\ \alpha} & \;
^{210}{\rm Pb}\; \substack{32.0\; y\\ \xrightarrow{\makebox[0.6cm]{}}\\ \beta} \; ^{210}{\rm Bi}\;
\substack{7.23\; d\\ \xrightarrow{\makebox[0.6cm]{}}\\ \beta} \; ^{210}{\rm Po}\;
\substack{200\; d\\ \xrightarrow{\makebox[0.6cm]{}}\\ \alpha} \; ^{206}{\rm Pb}\; \label{decay_chain}
\end{eqnarray*}

$(\alpha, n)$-reaction yields can be modelled reliably with codes like GEANT4~\cite{AGOSTINELLI2003}, SOURCES-4C~\cite{S4C_2002} and others, as shown in reference~\cite{geant4_an_2020}. What is less understood in the nuclear and particle physics literature is what fraction of the radon progeny $^{218}$Po, $^{214}$Pb and $^{214}$Bi in the air attach to surfaces. 
A wide range of attachment lengths can be found in the literature~\cite{leung_2005,guiseppe_2011,stein_2018,morrison_2018}.
Results can be difficult to compare because of differing reporting units. It is further not known whether all or only some of the unstable radon progeny attach to surfaces.

Planning for parts handling and the need for costly radon removal devices requires knowledge of radon progeny attachment rates, coupled with experiment-dependent computed $^{210}$Po-activity requirements. These data can be used to formulate maximal exposure durations for experiment components to a specific radon environment. 
In this paper we present a systematic study of radon progeny attachment to surfaces and offer an interpretation of the results, challenging commonly held notions.
While the particular choice of studied materials was motivated by the needs of the nEXO experiment~\cite{nEXO:2022nam}, the scope of the study is general, its results broadly relevant to low background experiments.

\subsection{Radon and its Progeny in Air}\label{sec:intro}
As a member of the $^{238}$U decay chain, found in soil, the emanated radioactive noble gas $^{222}$Rn is contained in ambient air. 
The global geometric mean of the outdoor volumetric activity is $\mathcal{A}_{Rn}=45\; \rm Bq/m^3$~\cite{UNSCEAR_2008}. 
Depending on the ventilation, indoor concentrations can be 100 Bq/m$^3$ or more. The SNOLAB Technical Reference Manual reports an average $^{222}$Rn specific activity of $\mathcal{A}_{Rn}=(123.2\pm 13.0)$ Bq/m$^3$ for the underground lab~\cite{snolab_handbook_2016}. Because of its chemical inactivity and relatively long lifetime, radon is typically distributed homogeneously in room air.

Its progeny $^{218}$Po, $^{214}$Pb and $^{214}$Bi, chemically active heavy metals, some electrically charged, show a much more complex behaviour.
They participate in one or more of the following processes: neutralization, chemical reaction, including hydratization, attachment to aerosol particles and attachment to surfaces. 

For the attachment of radon progeny to surfaces we assume: only the radon progeny $^{218}$Po, $^{214}$Pb and $^{214}$Bi are being collected. $^{214}$Po ($\rm T_{1/2}=163.6\; \mu s$) decays instantaneously together with its progenitor $^{214}$Bi, resulting in equal decay rates. 
Every $^{214}$Po decay results in the creation of a $^{210}$Pb atom. The activity concentration of $^{210}$Pb in air is known to be low, due to its long half-life that prevents it from decaying while airborne~\cite{daish_2005}. Its direct collection plays no role.


Previous attachment studies reported in~\cite{leung_2005,guiseppe_2011,stein_2018,morrison_2018,nastasi_2019} assumed that radon and its progeny, because of their relatively short half-lives, are in secular equilibrium in air. However, this is not the case since decay is not the only removal process for these chemically active species~\cite{nazaroff_nero_1988}. 
The resulting chain imbalance introduces environment-dependent variability. The study presented here does not assume sub-chain equilibrium. 

Radon progeny equilibrium is a well studied subject because it matters for the estimation of human radiation exposure.
Departure from chain equilibrium is often described via equilibrium factors
$f_{i}=\frac{\mathcal{A}_{i}}{\mathcal{A}_{Rn}}$, with $\mathcal{A}_{Rn}$ and $\mathcal{A}_{i}$ denoting the volumetric activities of $^{222}$Rn, $^{218}$Po, $^{214}$Pb, and $^{214}$Bi in air, respectively (e.g. in units of Bq/m$^3$). The NRC~\cite{nrc_2011} gives typical $^{222}$Rn: $^{218}$Po : $^{214}$Pb : $^{214}$Bi activity ratios of 1 : 0.5 : 0.3 : 0.2 for indoor environments.
Extensive literature exists on a particular linear combination of the $f_{i}$-values and its time and environment dependence. Literature on individual $f_i$ values could not be found.

\subsection{Radon Progeny Attachment to Surfaces}
\label{sec:intro_attachment}
During air exposure, radon progenies get attached (gained) to a surface and lost due to their radioactive decay. 
Let the number density of the atoms of species $i$ in the air available for attachment to surfaces be $\mathcal{C}_i$ (in units of atoms/m$^3$). The instantaneous capture rate $g_i$ of atoms of species $i$ on a surface $S$ is: $g_i=v_i\cdot S\cdot \mathcal{C}_i$. The constant of proportionality $v_i$ has dimension of speed. Alternatively, the gain term may be expressed through the volumetric activity $\mathcal{A}_i$ (in units of Bq/m$^3$) of the collected species: $g_i=v_i\cdot S\cdot \mathcal{A}_i\cdot \tau_i$. Using this metric, the constant of proportionality $d_i=v_i\cdot \tau_i$ has dimension of length. Therefore, collection lengths and collection speeds are equivalent, related to each other via the mean lifetime $\tau_i$.
Finally, when expressing the source strength in terms of the $^{222}$Rn volumetric activity one gets: $g_i=d_i\cdot S\cdot f_i\cdot \mathcal{A}_{Rn}$, introducing chain equilibrium. 

The number of $^{218}$Po, $^{214}$Pb and $^{214}$Bi atoms, $N_{Po}(t_d)$, $N_{Pb}(t_d)$, and $N_{Bi}(t_d)$, present on a surface $S$ at (deposition) time $t_d$ is given by:
{\footnotesize
\begin{eqnarray}
\frac{dN_{Po}(t_d)}{dt_d} & = & S\cdot f_{Po}\cdot d_{Po} \cdot \mathcal{A}_{Rn} - \frac{N_{Po}(t_d)}{\tau_{Po}} \label{growth_218po}
\\
\frac{dN_{Pb}(t_d)}{dt_d} & = & S\cdot f_{Pb}\cdot d_{Pb} \cdot \mathcal{A}_{Rn} + \frac{N_{Po}(t_d)}{\tau_{Po}} - \frac{N_{Pb}(t_d)}{\tau_{Pb}} 
\label{growth_214pb}
\\
\frac{dN_{Bi}(t_d)}{dt_d} & = & S\cdot f_{Bi}\cdot d_{Bi} \cdot \mathcal{A}_{Rn} + \frac{N_{Pb}(t_d)}{\tau_{Pb}} - \frac{N_{Bi}(t_d)}{\tau_{Bi}} 
\label{growth_214bi}
%
\end{eqnarray}
}
%
The $f_i\cdot d_i$-values measure the degree of radon progeny attachment to a given surface, they are the quantities observed in this study. We call these products the effective equilibrium-dependent collection distances.

As discussed in reference~\cite{nazaroff_nero_1988}, $^{214}$Pb, born in $\alpha$-decays, carries recoil energy and can desorb from aerosols. It was found that the corresponding loss term is degenerate with the effective collection length. It cannot be determined unambiguously from the data. We, therefore, absorb this term into the collection length.

In steady state ($\frac{dN_{i}(t_d)}{dt_d}\approx0$, for exposure durations $t_d\gg \tau_i$) the $^{218}$Po, $^{214}$Pb and $^{214}$Bi activities $A_i(t_d)=\frac{N_i(t_d)}{\tau_i}$ on $S$ are:
{\footnotesize
\begin{eqnarray}
\frac{A_{Po}(t_d)}{S} & \approx & \mathcal{A}_{Rn} \cdot f_{Po}\cdot d_{Po} 
\label{eq:po_growth_approx}
\\
\frac{A_{Pb}(t_d)}{S} & \approx & \mathcal{A}_{Rn} \cdot \left(f_{Po}\cdot d_{Po} + f_{Pb}\cdot d_{Pb}\right) 
\label{eq:pb_growth_approx}
\\
\frac{A_{Bi}(t_d)}{S}  & \approx & \mathcal{A}_{Rn} \cdot \left(f_{Po}\cdot d_{Po} + f_{Pb}\cdot d_{Pb}+ f_{Bi}\cdot d_{Bi}\right) \label{eq:bi_growth_approx}
%
\end{eqnarray}
}
Comparison to the full time-dependent solutions shows less than 5\% deviation of the approximate solutions for exposure times in excess of 3 hours. The effective collection lengths are, therefore, given by the ratios of the steady state surface activities over the specific $^{222}$Rn activity of the air.

Since the  $^{214}$Bi decay rate and the $^{210}$Pb growth rate are equal, the $^{210}$Pb growth rate $R_{210Pb}$ (here in atoms deposited per unit time) is determined by the environmental radon activity and the steady state $^{214}$Bi activity for long exposures (longer than 3 h):
\begin{equation}
    \frac{R_{210Pb}}{S}=\mathcal{A}_{Rn} \cdot \left(f_{Po}\cdot d_{Po} + f_{Pb}\cdot d_{Pb}+ f_{Bi}\cdot d_{Bi}\right)
    \label{eq:210pb_growth}
\end{equation}
The time and environment dependence of the equilibrium factors $f_i$ are a source of variability. Collection distances determined in one location have to be translated to the conditions at a different environment using modelling. The summed effective collection length $d_{sum}=f_{Po}\cdot d_{Po} + f_{Pb}\cdot d_{Pb}+ f_{Bi}\cdot d_{Bi}$ alone determines the steady state growth rate of $^{210}$Pb.

\subsection{Radon Progeny Decay} 
\label{sec:intro_decay}
Compared to a direct measurement of the $^{210}$Pb growth via $^{210}$Po $\alpha$-decays, the observation of the decay of its short-lived progenitors offers several advantages: fast sample turn-around, low background rates and higher specific activities. The main disadvantage of the chosen method is the somewhat cumbersome mathematics, discussed in this paper.

Observation of the time dependence of  $^{218}$Po and $^{214}$Po $\alpha$-decays, after ending the contact with air, allows to determine $d_{sum}$. 
This is done by exposing samples of various materials to air to accumulate activities.
Shortly after the end of exposure the samples are transferred into a vacuum chamber for counting. 
During the decay phase the rate of change of the different species is given by equations~\ref{growth_218po}, \ref{growth_214pb} and \ref{growth_214bi} with the gain terms set to zero: $d_i=0$.

As boundary conditions it is assumed that at time $t=0$ of the decay phase, the surface activities, given by equations~\ref{eq:po_growth_approx}, \ref{eq:pb_growth_approx} and \ref{eq:bi_growth_approx} for an exposure duration $t_d$, are present. The solutions of these equations are the time-dependent activities $A_i(t)$:
\begin{widetext}
\begin{eqnarray}
\frac{A_{Po}(t)}{S} & = & \mathcal{A}_{Rn} \cdot  (f_{Po}\cdot d_{Po})\cdot e^{-t/\tau_{Po}} \label{eq:po_decay} \\
                        &     &  \nonumber \\
                        &     &  \nonumber \\
\frac{A_{Pb}(t)}{S} & = & \mathcal{A}_{Rn} \cdot \left[ \left(f_{Po}\cdot d_{Po} + f_{Pb}\cdot d_{Pb}\right)\cdot e^{-t/\tau_{Pb}} - (f_{Po}\cdot d_{Po})\cdot \frac{\tau_{Po}}{\tau_{Po}-\tau_{Pb}}\cdot \left( e^{-t/\tau_{Pb}} - e^{-t/\tau_{Po}}  \right)  \right] \label{eq:pb_decay}\\
                                   &    &  \nonumber \\
                                   &     &  \nonumber \\
 \frac{A_{Bi}(t)}{S} & = & \mathcal{A}_{Rn} \cdot \left[   \left(f_{Po}\cdot d_{Po} + f_{Pb}\cdot d_{Pb}+ f_{Bi}\cdot d_{Bi}\right)  \cdot e^{-t/\tau_{Bi}} +      \right.  \nonumber \\
                                   &    &   \left(f_{Po}\cdot d_{Po} + f_{Pb}\cdot d_{Pb}\right) \cdot  \frac{\tau_{Pb}}{\tau_{Pb}-\tau_{Bi}}\cdot \left( e^{-t/\tau_{Pb}} - e^{-t/\tau_{Bi}}  \right) +  \nonumber \\
                                   &    & \left. (f_{Po}\cdot d_{Po})\cdot  \frac{\tau_{Po}}{\tau_{Po}-\tau_{Pb}}\cdot \left\{ \frac{\tau_{Pb}}{\tau_{Pb}-\tau_{Bi}}\cdot \left( e^{-t/\tau_{Bi}} - e^{-t/\tau_{Pb}}  \right)  - \frac{\tau_{Po}}{\tau_{Po}-\tau_{Bi}}\cdot \left( e^{-t/\tau_{Bi}} - e^{-t/\tau_{Po}}  \right)   \right\} \right] \label{eq:bi_decay}           
\end{eqnarray}
\end{widetext}
The energy of the $\alpha$-peaks separates $^{218}$Po from $^{214}$Po decays.
Fits to the time dependence of the $\alpha$-activities $A_{Po}(t)$ and $A_{Bi}(t)$ can be used to derive the effective collection lengths (equivalent to the activities at time zero). A simple exponential fit of the time dependence of the $^{218}$Po peak to equation~\ref{eq:po_decay} gives $f_{Po}\cdot d_{Po}$ and its uncertainty. The time-dependent activity of the $\beta$-decaying $^{214}$Pb is not determined using $\alpha$-spectroscopy. 
$\left( f_{Po}\cdot d_{Po} + f_{Pb}\cdot d_{Pb}\right)$ and $\left( f_{Po}\cdot d_{Po} + f_{Pb}\cdot d_{Pb}+ f_{Bi}\cdot d_{Bi}\right)$ are obtained from a time fit to $A_{Bi}(t)$ using equation~\ref{eq:bi_decay}. $f_{Po}\cdot d_{Po}$, obtained from the $^{218}$Po decay data, is used in that fit as initial value and is allowed to float, constrained by its uncertainty.
$A_{Bi}(t)$ has a non-trivial time dependence, depending on the unknown effective collection lengths. Its shape is non-exponential.

Note that the time integral of equation~\ref{eq:bi_decay}, equivalent of the number of events under the $^{214}$Po $\alpha$-peak when taken from an initial time of 0 to a final time of $\infty$, is proportional to the sum of all initial activities. It is, thus, not a good measure of $d_{sum}$, needed to determine steady state $^{210}$Pb production.

\begin{figure}[bt]
\includegraphics[width=8.5cm]{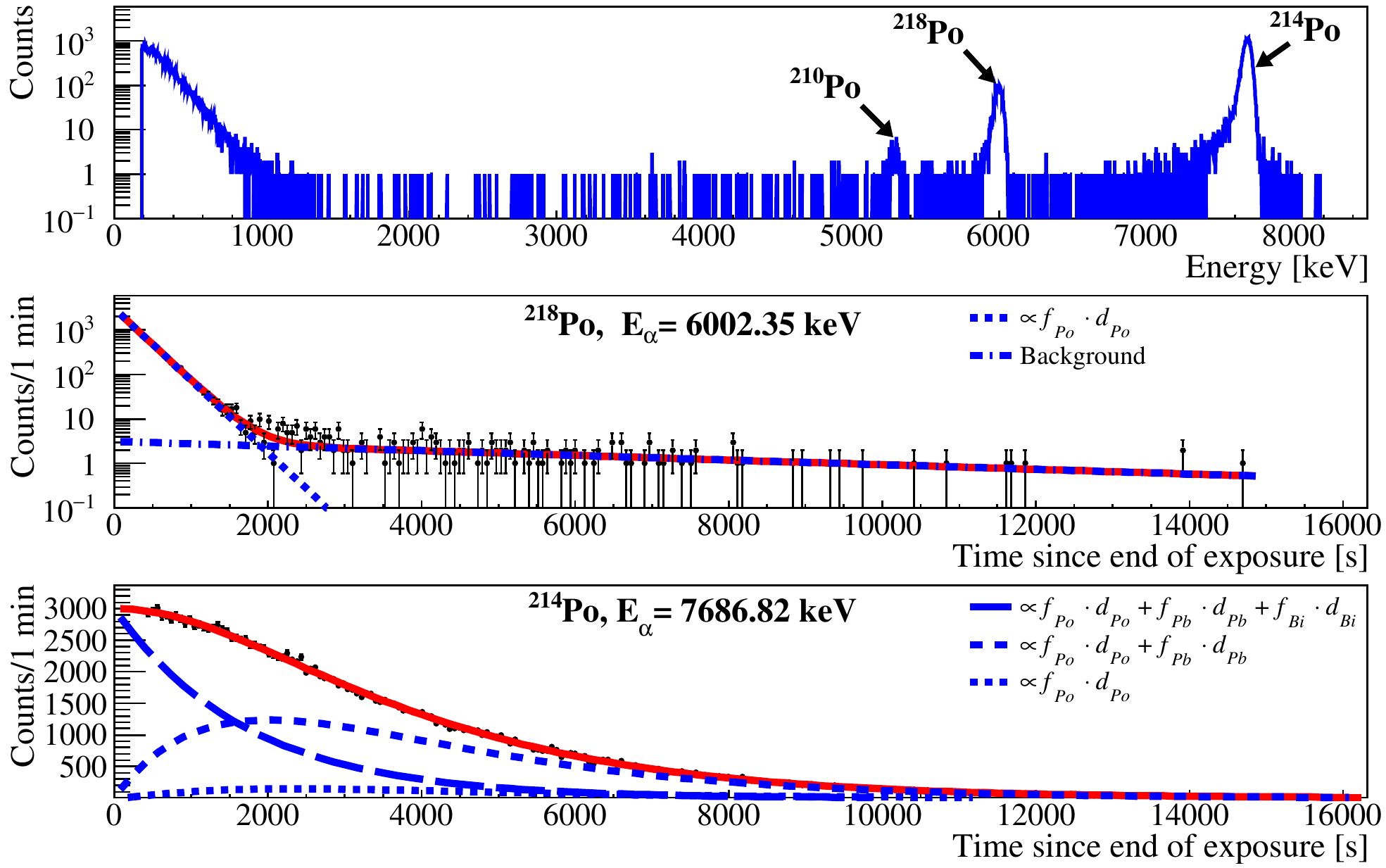}
\caption{\label{fig:time_curves} 
$\alpha$-energy spectrum and time dependences obtained with a copper sample exposed to a high-activity $^{222}$Rn source. This high statistics data is used to verify the multi-component time fit for $^{214}$Po events. 
Top panel: $\alpha$-energy spectrum. Because of their short range, the $\alpha$-particles have to originate from the surface of the sample. The $^{214}$Po and $^{218}$Po peaks are resolved. The lower energy $^{218}$Po peak is superimposed to a small tail of $^{214}$Po events. 
Middle panel: time dependence of the $^{218}$Po peak. The small-dotted line is the exponential fit to the $^{218}$Po decay, the dash dotted line shows the time-dependent background due to $^{214}$Po tail events.
Bottom panel: time dependence of the $^{214}$Po peak. The solid line shows the fit to the data, using equation~\ref{eq:bi_decay}. $\chi^2/ndf=230.6/235$ indicates good fit quality. The terms proportional to $f_{Po}\cdot d_{Po}$ (the $^{218}$Po starting activity) in equation~\ref{eq:bi_decay} are shown by the small-dotted line, terms proportional to $f_{Po}\cdot d_{Po}+f_{Pb}\cdot d_{Pb}$ (the $^{214}$Pb starting activity) by the short-dashed line and those proportional to $f_{Po}\cdot d_{Po}+f_{Pb}\cdot d_{Pb}+f_{Bi}\cdot d_{Bi}$ (the $^{214}$Bi starting activity) by the dashed line.
}
\end{figure}

The validity of the decay model has been verified by exposing surfaces to a Pylon RN-1025 $^{222}$Rn source in a sealed exposure box. Example data is shown in figure~\ref{fig:time_curves}. 
Note that the observed time dependence is consistent with the presence of radon progeny only, no long-lived  $^{222}$Rn component with T$_{1/2}=3.822$ d is observed. We conclude that there is negligible attachment of $^{222}$Rn or diffusion into the sample.
Equations~\ref{eq:po_decay} and \ref{eq:bi_decay} result in good fits to the high statistics data. 

\section{Measurement of Radon Progeny decays}
For this study, samples made of copper, Hamamatsu SiPM (silicon photomultiplier), high density polyethylene (HDPE), carbon fiber composite, sapphire, nickel, metallized silica, fused silica, and polytetrafluoroethylene (PTFE) were tested. The attachment of radon progeny has been measured using, in most cases, circular disks with a diameter ranging from 61.4 to 76.2 mm and 1.6 to 3.2 mm thickness. The two Hamamatsu-provided SiPM samples had rectangular shape with a side length of about 67 mm and a thickness of 0.63 mm.
To account for these shape differences, the detector acceptance was calculated for all samples using a GEANT4 simulation, as described below. 
With the exception of the SiPM, all sample disks had smooth surfaces. The SiPM sample has a unique surface structure, as required by its functionality.

\subsection{The Experiment Setup}
The samples were exposed to air in an unventilated basement in Gallalee Hall at the University of Alabama (UA).
The ambient $^{222}$Rn activities sometimes reached $400\; \rm Bq/m^3$. A soft-wall mini-clean room ($\rm{1.1\; m \times 1.5\; m\times 2.1\; m}$) was erected to control the dust content of the air and to remove radon progeny by means of HEPA filtration. The dust content of the air was monitored using a ParticleScan CR airborne particle counter. The $^{222}$Rn-concentration was continuously monitored with a Durridge Rad7 electronic radon detector. For non-conducting samples, the surface potential was measured with a model EFM115 Transforming Technologies electrostatic field meter at the beginning and end of the exposure. The temperature and relative humidity were continuously monitored using a LogTag HAXO-8 humidity and temperature logger. The humidity level of the air could be controlled using a (de)humidifier. The atmospheric pressure was recorded with a Sun Nuclear Model 1029 continuous radon monitor. The air flow of the HEPA filtration was calibrated at its low and high settings using a TSI Alnor capture hood model EBT731. For part of the campaign a Bladewerx SabreBPM2 detector was used to monitor the concentration of radon progeny in the clean room air, with a focus on $^{218}$Po and $^{214}$Po $\alpha$-decays.
Surface attachment samples were exposed for a minimum of 3 hours to assure the $^{218}$Po, $^{214}$Pb and $^{214}$Bi surface activities reached saturation.

Shortly after the end of exposure, the samples were transferred into a vacuum chamber for $\alpha$-counting. This transfer took on average (determined for the measurements with Cu samples) 1.3 minutes. About half of the delay time was spent pumping down the counting chamber.

The surface activities of the samples were determined using two ORTEC low background ULTRA ENS-U3000 Si detectors with $30\; \rm cm^2$ active area, operated in two separate vacuum chambers. Two ORTEC Alpha Mega integrated measurement systems are used. The energy scales and resolution characteristics of the devices are monitored with Eckert $\&$ Ziegler $^{210}$Po and mixed $\alpha$-sources, the latter containing $^{239}$Pu, $^{241}$Am, and $^{244}$Cm activities. Both sources have activity calibration certificates provided by the manufacturer. The detector's energy resolution is determined by fitting source data to a two-tailed Gauss function, as described in~\cite{alpha_peak_1987,alpha_peak_2015}. The resulting energy resolution (taken to be the parameter $\sigma$ in~\cite{alpha_peak_1987}) around 5300 keV is about 20 keV.

Because the study presented here required multiple measurements, the dependence of the $\alpha$-peak centroid and resolution on the chamber pressure was studied using the mixed source. No impact on the peak was found at chamber pressures of 500 mTorr and below, with the pressure determined by an external InstruTech CVM211 Stinger Pirani vacuum gauge. To assure data uniformity, sample measurements were started once a chamber pressure of 500 mTorr had been achieved. Oil free Leybold SCROLLVAC SC 15D and Pfeiffer Vacuum HiScroll 12 scroll pumps with a pump speed of more than $10\; \rm m^3/h$ were used to achieve quick pump-down because of the short half-life of $^{218}$Po.

\subsection{Detector Simulation}
%

%
In order to convert $\alpha$-peak derived counting rates into activities, a GEANT4 simulation code was developed. This code allows to calculate detection efficiencies for $\alpha$-particles uniformly deposited on a sample surface. 
The GEANT4 model was prepared using information provided by the detector manufacturer.

The validity of the simulation was verified by means of comparison to data obtained with the activity-calibrated $^{210}$Po $\alpha$-source.
The source activity at the time of measurement was 1.66 $\pm$ 0.02 Bq. The counting rate after 1 hour of measurement was estimated by integrating the 5.3 MeV $\alpha$-peak.
The measured detection efficiency, calculated as the ratio between the counting rate and source activity, was $(38.6 \pm 0.9)\%$. A detailed model of the $\alpha$-source was implemented into the GEANT4 code. The simulated source detection efficiency was $(38.57 \pm 0.06)\%$. It should be noted that both random and systematic uncertainties are reported for the measured detection efficiency. For the simulation only the statistical error is reported.
The agreement between the simulated and measured detection efficiencies validates the detector acceptance model.

\subsection{Data Analysis}
Data acquisition consists of an ORTEC Alpha Mega integrated 12-bit digitizer and MAESTRO software.
Data was collected in the form of histograms, saved every minute to allow a time-differential analysis. 
The energy analysis, distinguishing $^{218}$Po from $^{214}$Po decays, utilizes integration over a broad energy window. The low statistics of typical data runs did not allow to fit peaks with a tail. This approach has the advantage that it integrates tail and peak and is, thus, insensitive to the possible embedding of $^{218}$Po and $^{214}$Po into the sample, resulting in non-trivial peak to tail ratios.

To account for the low number of counts often observed in the $\alpha$-spectra, the fits are performed assuming Poissonian statistics plus Gaussian constraint terms. 
The surface activities and their uncertainties 
are obtained by minimizing the resulting negative logarithmic likelihood function.

The likelihood function is used to estimate the statistical uncertainty of each run. The variability of the derived effective deposition lengths was found to be substantial (between 20 and 54\%) for all studied materials. To get a better handle on the large variability of the effective deposition lengths, it was decided to perform repeated measurements for all configurations and determine the variance from the data. For most configurations about 25 repeated measurements were performed, for one copper configuration 83. For the latter configuration (copper in the condition ``fan off, tent open''), the average statistical error was found to be a factor 3.3 smaller than the variance-derived standard deviation. 

The determination of effective radon progeny collection lengths requires knowledge of the specific $^{222}$Rn activity of the air the samples are exposed to. The $^{222}$Rn volumetric activity was monitored over a two year period in the basement of Gallalee Hall at UA, using the RAD7 detector. An integration time of 2 h was chosen to be able to resolve time transients.
%
The surface activity data, determined by alpha counting, is correlated with the $^{222}$Rn by means of the time record. To account for the variability of the radon activity but still keep the data analysis at a reasonable level of complexity, a single activity value calculated at the time of sample extraction was assigned to each measurement.  
The radon activity was obtained by means of linear interpolation between the two nearest RAD7 measurements, bracketing the sample extraction time.
Time transients faster than the chosen two hour integration time of the RAD7 detector cannot be captured.
The statistical error of each radon measurement was derived from the instrument sensitivity, supplied by the manufacturer.

\begin{figure}[bt]
\includegraphics[width=8.5cm]{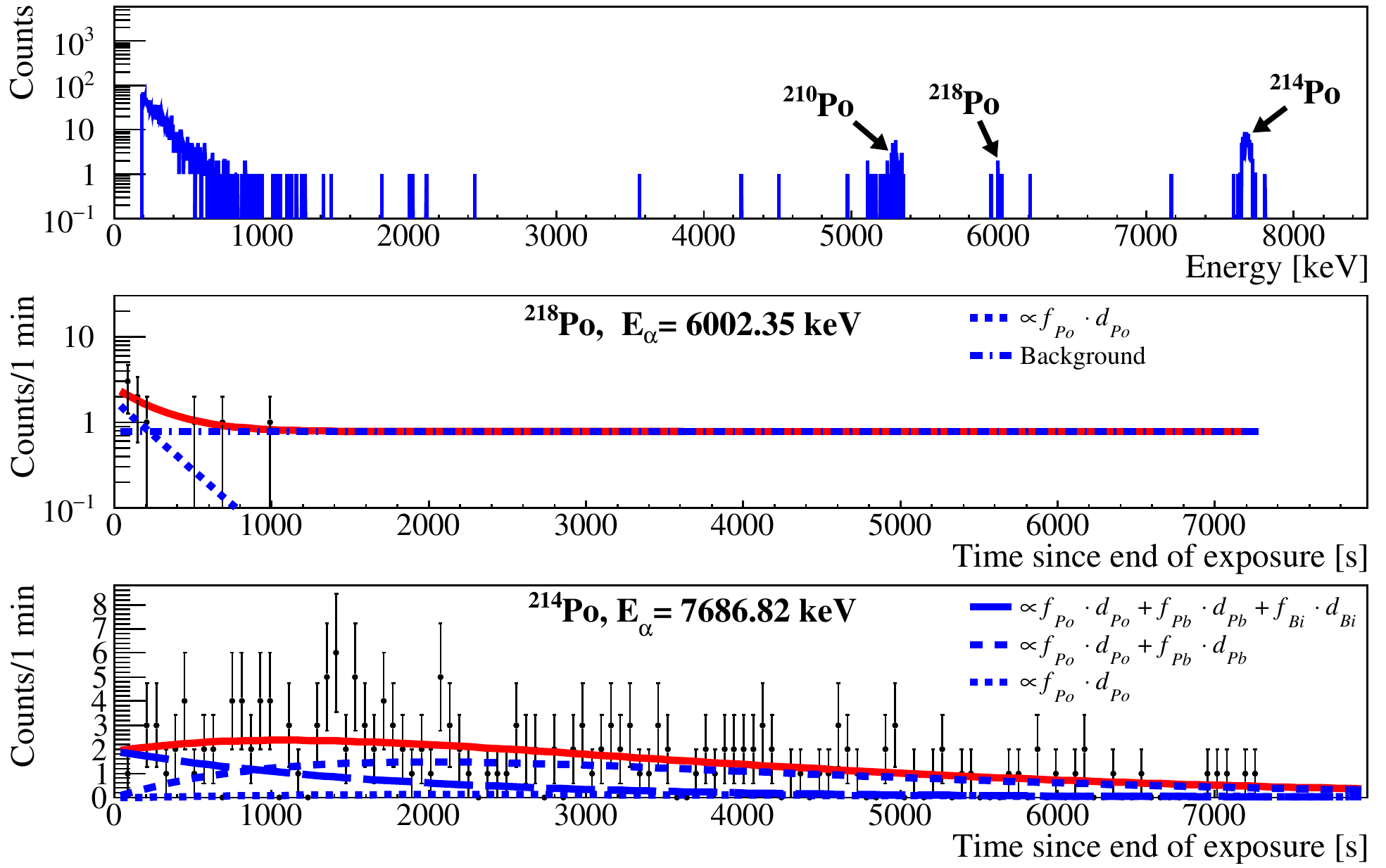}
\caption{\label{fig:basement_copper_example} 
Example data for an $\alpha$-energy spectrum (top) and time dependence of the $^{218}$Po (middle) and $^{214}$Po (bottom) $\alpha$-peak areas, for a 1 minute binning. This data was obtained with a copper sample exposed to laboratory air in ``fan off, tent closed'' condition. The fits to the time distributions are shown. Note that the $^{218}$Po population for this run is estimated at 24 atoms at the end of exposure. Due to the extremely low number of $^{218}$Po counts, the effective collection length $f_{Po}\cdot d_{Po}$ was determined by integrating the number of counts over the first 30 minutes after exposure instead of a fit to the time distribution.}
\end{figure}
Data was collected at differing flow conditions and surface to volume ratios to test the attachment rates and $f_{Po}$ against the Jacobi model, described in the next section. Radon progeny attachment data was taken in four different conditions: 1) air flow off with clean room closed (``fan off, tent closed'', estimated surface to volume ratio 5.0 m$^{-1}$), 2) air flow off with clean room open (``fan off, tent open'', estimated surface to volume ratio 2.0 m$^{-1}$), 3) air flow on low with clean room closed (``fan on low, tent closed''), and 4) air flow on high with clean room closed (``fan on high, tent closed''). For the runs with the ventilation active, the air flow rates were measured as $1.3\; \rm m^3/min$ (``fan on low'') and $10\; \rm m^3/min$ (``fan on high''), respectively. The resulting rates of volume exchange were $0.36\; \rm min^{-1}$ and $2.9\; \rm min^{-1}$. 
The attachment measurements at the SNOLAB Ladder Lab, reported in~\cite{stein_2018}, were performed at an air exchange rate of $10\; \rm h^{-1} = 0.167\; \rm min^{-1}$ \cite{snolab_handbook_2006}. 
To the best of our knowledge, the estimated surface to volume ratio at SNOLAB was 0.8 m$^{-1}$ \cite{private_comm_jardin}
for the radon progeny attachment measurements reported in~\cite{stein_2018}. Because of our interest in translating our results to the conditions at SNOLAB, a large fraction of measurements was taken at the ``fan off, tent open'' and ``fan on low, tent closed'' conditions.

All attachment data were analyzed by means of the time fit, described above.
However, for data with a low number of events the effective collection length $f_{Po}\cdot d_{Po}$ was determined by integrating the number of $^{218}$Po counts over the first 30 minutes after exposure instead of a time fit.
Figure~\ref{fig:basement_copper_example} shows an example of a low event rate data set.

In addition, individual effective lengths were determined by means of subtracting the summed effective lengths entering equations~\ref{eq:po_decay}, \ref{eq:pb_decay} and \ref{eq:bi_decay} from each other, exposure run-by-run to cancel the effect of temporal variations. 
Results derived from repeated runs, done under a certain ventilation condition, were summarized by means of frequency distributions of these individual effective lengths, as shown in figures~\ref{fig:copper_lengths} and \ref{fig:all_fan_off_lengths}. Only runs corresponding to a sample surface potential of less than 100 V and ambient radon concentrations of at least $50\; \rm Bq/m^3$ were utilized.

\section{Results and Discussion}
The study presented here covers many aspects of this somewhat complex subject. To enhance readability, the following section is organized around a few simple questions for which we offer answers.


\subsection{What is being Collected?}
\begin{figure}[bt]
\includegraphics[width=8.5cm]{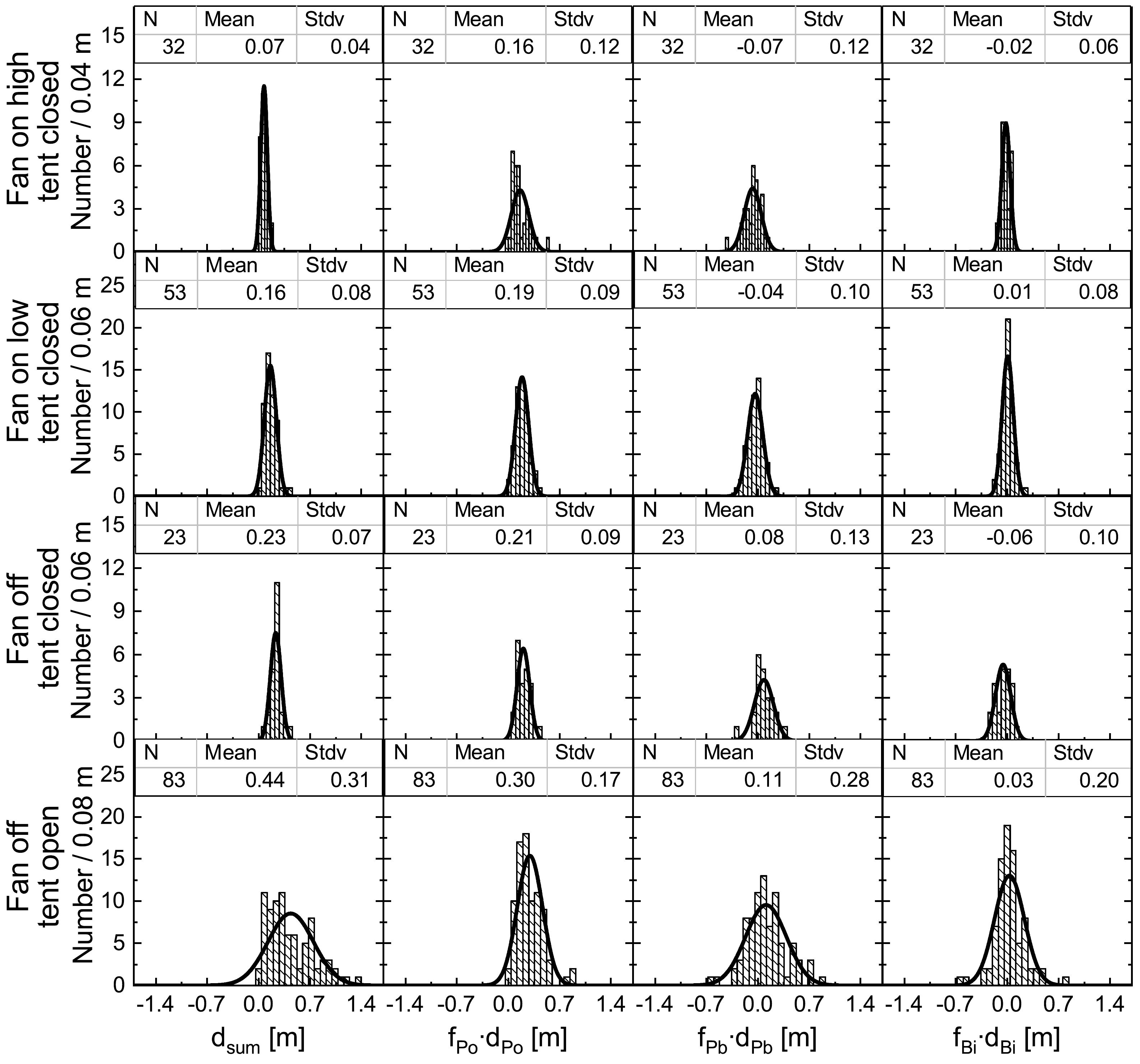}
\caption{\label{fig:copper_lengths} 
Frequency distributions of the summed and individual effective collection lengths observed for copper. The parameter $N$ denotes the number of attachment measurements entering into any particular histogram. Means and standard deviations are given.}
\end{figure}
As discussed in section~\ref{sec:intro}, the airborne radon progeny $^{218}$Po, $^{214}$Pb and $^{214}$Bi, in principle, can all attach to surfaces. In this section we present our evidence that collection is dominated by $^{218}$Po.

Equation~\ref{eq:bi_decay} describes the time dependence of the $^{214}$Po decay after end of contact with the environment. 
A fit to the time dependence, using equation~\ref{eq:bi_decay}, determines all three effective collection lengths.
$f_{Po}\cdot d_{Po}$ values and fit errors, derived from equation~\ref{eq:po_decay}, serve as constraint terms in equation~\ref{eq:bi_decay}, previously described in section~\ref{sec:intro_decay}. As final results we report fit values derived from equation~\ref{eq:bi_decay}.
Repeated measurements and fits to the time-dependent activities of radon progeny attachment determine the mean and standard deviations of the fit results, by means of frequency distributions.

The practical problem in interpreting the fit results is the rather large variability. To address this difficulty the fit results are visualized as frequency distributions in histograms like figure~\ref{fig:copper_lengths}.
The left column of figure~\ref{fig:copper_lengths} shows the frequency distributions of $d_{sum}$, determining the $^{210}$Pb production, 
for copper and for various exposure conditions. The substantial variability of the fit results is evident.

In order to quantify the contributions of $^{214}$Pb and $^{214}$Bi attachment to the observed $^{214}$Po decays, we subtract the summed effective lengths, resulting from each fit to equation~\ref{eq:bi_decay}, appropriately from each other and then histogram the resulting differences. This approach pairs results corresponding to the same environmental $f_i$-values.
The second, third and fourth columns of figure~\ref{fig:copper_lengths} show the resulting histograms of the nuclide-specific effective lengths.
Comparing $d_{sum}$ to $f_{Po}\cdot d_{Po}$ shows for all environmental conditions the dominance of $^{218}$Po collection. At the highest ventilation rate, resulting in a low number of events (small $d_{sum}$), the differences even tend to be unphysical.

The same analyses were performed for the other materials, the resulting histograms are not shown. The same observation is made for all materials: $^{218}$Po provides the dominant contribution to the surface collection of radon progeny.

High statistics collection measurements, performed with the Pylon radon source and samples contained in a small, sealed exposure box give the same results: $^{218}$Po collection is dominant.

As discussed below, the Jacobi model indicates small equilibrium factors $f_{Pb}$ and $f_{Bi}$ at high rates of ventilation, offering an explanation for the smallness of the effective collection lengths $f_{Pb}\cdot d_{Pb}$ and $f_{Bi}\cdot d_{Bi}$. Collection might be small because of the impact of chain disequilibrium.

Although it is understood that our measurements are mainly sensitive to $^{218}$Po collection, we are reporting $d_{sum}$ values as our primary result. Its values are most closely related to the growth of $^{210}$Pb on surfaces, it therefore has the greatest utility.

\subsection{How much is being Collected?}
After understanding what radionuclide is mainly collected on surfaces the next question is how much of it?

 Because $d_{sum}$ is given by a sum it cannot directly be converted into an effective collection speed. This conversion can only be done for the individual, nuclide specific effective lengths, discussed in the previous section. A similar argument applies to an interpretation of the lengths as diffusion lengths, given as the square root of a diffusion constant times the mean lifetime. As shown in the previous section, the dominant contribution to $d_{sum}$ is from $^{218}$Po collection which can be converted into a collection speed, helping in the comparison to literature values.


\paragraph{Mini-clean room results}

\begin{figure}[bt]
\includegraphics[width=8.5cm]{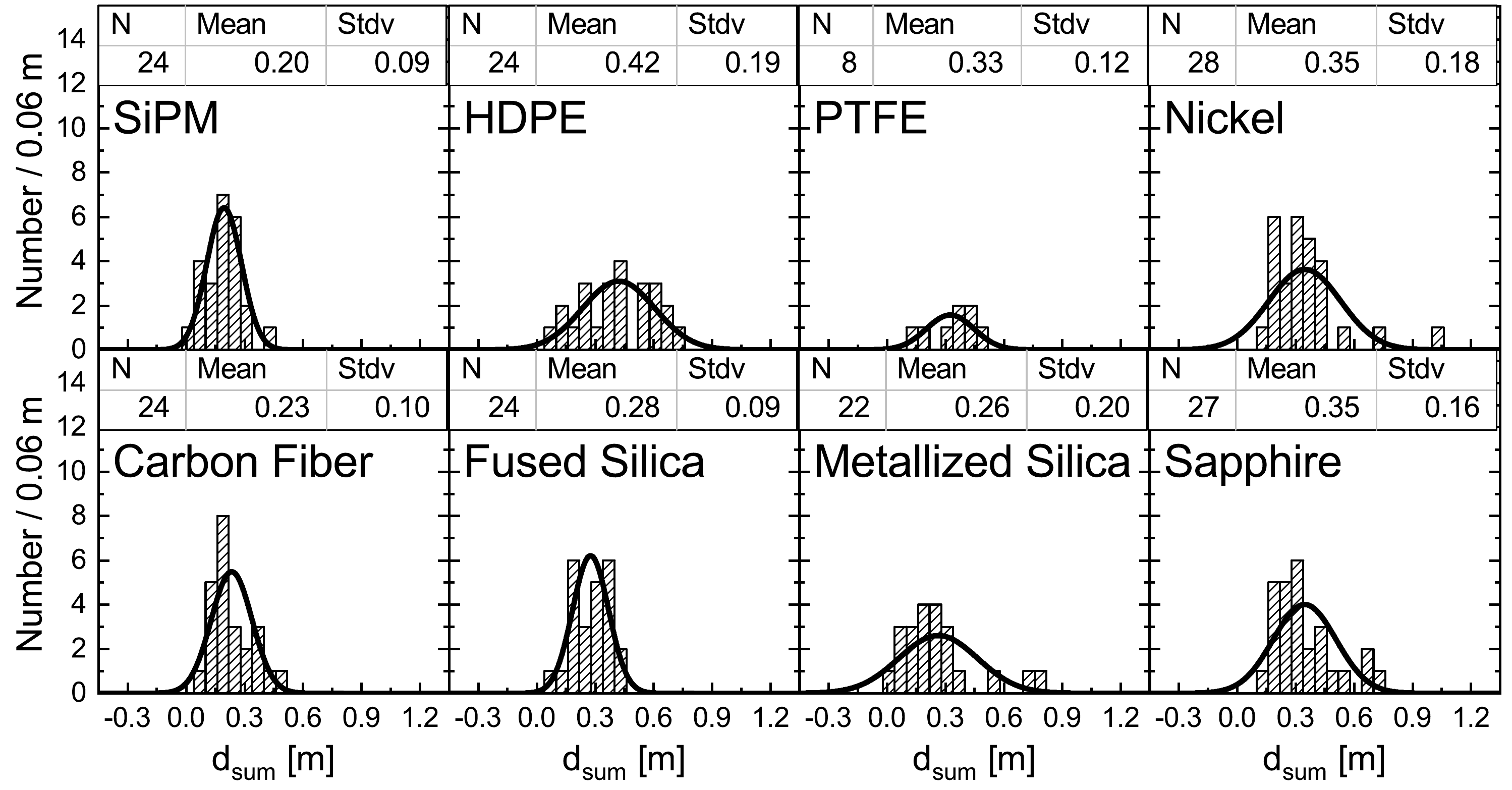}
\caption{\label{fig:all_fan_off_lengths} 
Frequency distributions of the summed effective collection lengths observed for Hamamatsu SiPM, HDPE, PTFE, nickel, carbon fiber composite, fused silica, metallized silica, and sapphire. The histograms are for the ``fan off, tent open'' condition.}
\end{figure}
Quantitative collection results, in form of $d_{sum}$ distributions resulting from repeated counting of Hamamatsu SiPM, HDPE, PTFE, nickel, carbon fiber composite, fused silica, metallized silica, and sapphire samples are shown in figure~\ref{fig:all_fan_off_lengths}. These measurements were performed in the “fan off, tent open” condition.

As seen in figures \ref{fig:copper_lengths} and \ref{fig:all_fan_off_lengths} the results show rather large dispersion. The material to material variability of the mean $d_{sum}$ values, on the other hand, is relatively small.

\begin{figure}[bt]
\includegraphics[width=8.5cm]{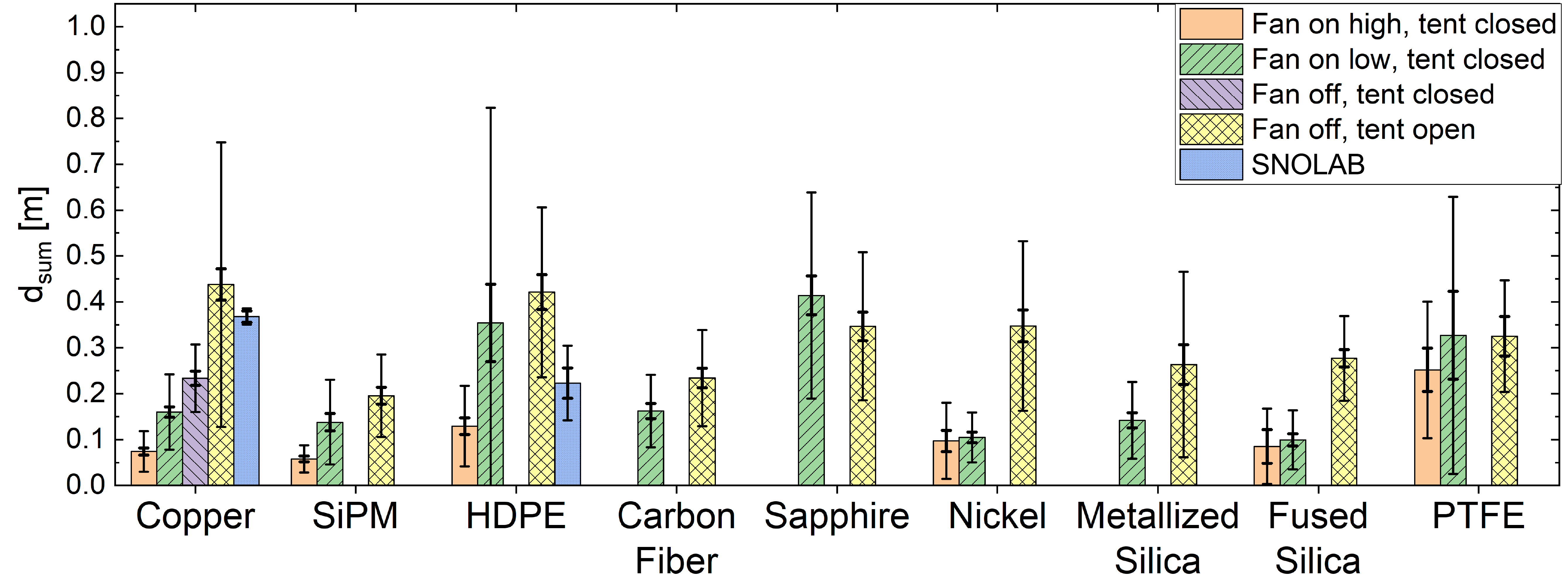}
\caption{\label{fig:all_lengths} 
Average summed effective collection lengths observed for all studied materials under all exposure conditions. Standard deviations and standard errors are reported with thin and thick error bars, respectively. The SNOLAB results obtained in \cite{stein_2018} with copper and HDPE samples are shown for comparison. We show their average lengths, standard deviations and standard errors, as calculated from table 4 in \cite{stein_2018}. \cite{stein_2018} reports only two measurements for copper and six for HDPE. The limited copper sampling is likely the reason for the small variability.
}
\end{figure}

Figure~\ref{fig:all_lengths} summarizes the average $d_{sum}$ values, their standard deviations and standard errors for all materials and all exposure conditions. A number of observations can be made.
The presence of ventilation and HEPA filtration reduces radon progeny attachment. This is a pronounced effect. The higher the air exchange rate the smaller $d_{sum}$.  Sapphire and PTFE seem to be exceptions but the large variability doesn't allow a clear statement. In the next section we explain this general trend with the Jacobi model, giving small equilibrium factors for large ventilation rates, in turn resulting in small effective attachment lengths.

For equal environmental conditions the $d_{sum}$-averages show rather limited material dependence.
As stated above, bias by the presence of electrical surface potentials was avoided by appropriate data selection.

In comparison to SNOLAB attachment results in~\cite{stein_2018}, our copper results for the ``fan off, tent open'' condition are a good match. The HDPE results don't agree that well. They show a difference with $3\sigma$ significance when using standard errors.

As mentioned above, the variability of the results in many cases exceeds the statistical error estimate.
We interpret this to indicate the presence of unaccounted  temporal variations impacting the results.
The time variability of the equilibrium factors, as a component of the effective deposition lengths, is one example.
Another source of variability lies in the chosen mathematical ansatz. Equations~\ref{growth_218po}, \ref{growth_214pb} and \ref{growth_214bi} treat the environmental radon activity as constant in time. However, the $^{222}$Rn activity fluctuates, and in the most general approach, should be treated as a time dependent function. To test the importance of the
latter, 24 attachment measurements with copper were performed back-to-back. Comparing the average $f_{Po}\cdot d_{Po}$ and its standard deviation derived from all runs, to those obtained when only analyzing 18 runs showing $^{222}$Rn time transients below 20\%, showed no significant difference. From this observation we conclude that our simplified mathematical treatment is, at least within the errors of this study, warranted. It should be noted that even the addition of $\mathcal{A}_{Rn}\rightarrow \mathcal{A}_{Rn}(t)=m\cdot t+b$ (obtained by linear interpolation between radon measurements) would result in a substantial complication of the math.

Our lab results are inconsistent with the rather short lengths reported in~\cite{guiseppe_2011} (after a suitable unit conversion). However, as shown below, much shorter lengths were observed by us too when using a radon source and small exposure box.


\begin{table}[htb]
    \centering
    \begin{tabular}{l|l|l|l}
        Sample & 
        Condition   & 
        $d_{sum}$ [m] &
        $f_{Po} \cdot d_{Po}$ [m] \\ \hline\hline
        Copper & 
        Rn source    &
        $0.0201 \pm 0.0022$ &
        $0.0203 \pm 0.0019$ \\
         & 
        fONH, tC    &
        $0.074 \pm 0.008$ &
        $0.161 \pm 0.021$ \\
         & 
        fONL, tC    &
        $0.160 \pm 0.011$ &
        $0.189 \pm 0.012$ \\
         & 
        fOFF, tC    &
        $0.234 \pm 0.015$ &
        $0.206 \pm 0.018$ \\
         & 
        fOFF, tO    &
        $0.438 \pm 0.034$ &
        $0.295 \pm 0.019$ \\ \hline
        SiPM  &
        Rn source    &
        $0.0278 \pm 0.0033$ &
        $0.0259 \pm 0.0030$ \\
         & 
        fONH, tC    &
        $0.058 \pm 0.007$ &
        $0.153 \pm 0.022$ \\
         & 
        fONL, tC    &
        $0.138 \pm 0.019$ &
        $0.276 \pm 0.041$ \\
         & 
        fOFF, tO    &
        $0.195 \pm 0.018$ &
        $0.181 \pm 0.026$ \\ \hline
        HDPE  &
        Rn source    &
        $0.0163 \pm 0.0022$ &
        $0.0156 \pm 0.0010$ \\
         & 
        fONH, tC    &
        $0.129 \pm 0.018$ &
        $0.468 \pm 0.068$ \\
         & 
        fONL, tC    &
        $0.354 \pm 0.084$ &
        $0.985 \pm 0.101$ \\
         & 
        fOFF, tO    &
        $0.421 \pm 0.038$ &
        $0.720 \pm 0.086$ \\ \hline
        Carbon &
        fONL, tC    &
        $0.162 \pm 0.016$ &
        $0.231 \pm 0.025$ \\
        Fiber & 
        fOFF, tO    &
        $0.234 \pm 0.021$ &
        $0.279 \pm 0.036$ \\ \hline
        Sapphire  &
        fONL, tC    &
        $0.414 \pm 0.043$ &
        $0.304 \pm 0.027$ \\
         & 
        fOFF, tO    &
        $0.347 \pm 0.031$ &
        $0.199 \pm 0.019$ \\ \hline
        Nickel  &
        fONH, tC    &
        $0.097 \pm 0.023$ &
        $0.286 \pm 0.065$ \\
         & 
        fONL, tC    &
        $0.105 \pm 0.011$ &
        $0.156 \pm 0.018$ \\
         & 
        fOFF, tO    &
        $0.348 \pm 0.035$ &
        $0.281 \pm 0.035$ \\ \hline
        Metallized  &
        fONL, tC    &
        $0.142 \pm 0.016$ &
        $0.203 \pm 0.023$ \\
        Silica & 
        fOFF, tO    &
        $0.263 \pm 0.043$ &
        $0.178 \pm 0.022$ \\ \hline
        Silica  &
        fONH, tC    &
        $0.085 \pm 0.037$ &
        $0.196 \pm 0.063$ \\
         & 
        fONL, tC    &
        $0.099 \pm 0.013$ &
        $0.171 \pm 0.021$ \\
         & 
        fOFF, tO    &
        $0.277 \pm 0.019$ &
        $0.232 \pm 0.035$ \\ \hline
        PTFE  &
        Rn source    &
        $0.0157 \pm 0.0010$ &
        $0.0172 \pm 0.0015$ \\
         & 
        fONH, tC    &
        $0.252 \pm 0.047$ &
        $1.187 \pm 0.179$ \\
         & 
        fONL, tC    &
        $0.327 \pm 0.096$ &
        $0.508 \pm 0.062$ \\
         & 
        fOFF, tO    &
        $0.325 \pm 0.043$ &
        $0.991 \pm 0.234$ \\ \hline
    \end{tabular}
    \caption{Mean $d_{sum}$ and $f_{Po}\cdot d_{Po}$ effective collection lengths obtained for all studied materials and all exposure conditions. The standard errors of the means are reported. ``fONH, tC'' stands for ``fan on high, tent closed'', ``fONL, tC'' $-$ ``fan on low, tent closed'', ``fOFF, tC'' $-$ ``fan off, tent closed'', and ``fOFF, tO'' $-$ ``fan off, tent open''. The estimated surface to volume ratio for the ``tent open'' condition was 2.0 m$^{-1}$, for ``tent closed'' 5.0 m$^{-1}$, for the Rn source measurements 55.2 m$^{-1}$. The rate of air exchanges is estimated as 0.36 min$^{-1}$ for the ``fan on low'' and 2.9 min$^{-1}$ for the ``fan on high'' condition. Reference~\cite{snolab_handbook_2006} gives an air exchange rate of 0.17 min$^{-1}$ for SNOLAB. The surface to volume ratio at SNOLAB is estimated as 0.8 m$^{-1}$ \cite{private_comm_jardin}.}
    \label{tab:all_length}
\end{table}

To make the numerical data more accessible, the measured effective collections lengths are summarized in table~\ref{tab:all_length}. Standard errors are stated.

Comparing $f_{Po}\cdot d_{Po}$ and $d_{sum}$-values reported in table~\ref{tab:all_length}, one can
see that the former often significantly exceeds the latter. One can call this behaviour unphysical. This is obvious for runs with ``fan on'' where disequilibrium is most pronounced. Our mathematical model contains no treatment of the time-dependence of the equilibrium factors. This means that statistical fluctuations could contribute to this behaviour. Furthermore, the $^{220}$Rn progeny $^{212}$Bi is not energy-resolved from $^{218}$Po alphas. This may lead to an overestimation of the $^{218}$Po collection, serving as initial guidance and external constraint (see section~\ref{sec:intro_decay}) for the $^{214}$Bi time fit. While this reasoning is plausible, we have no quantitative explanation. However, it is $d_{sum}$ that matters for $^{210}$Pb production. Its value is well determined via the time fit of the $^{214}$Po $\alpha$-peak. It is this direct connection to $^{210}$Pb growth that leads us to interpret $d_{sum}$ as the principal observable and not $f_{Po}\cdot d_{Po}$ having no internal consistency checks. Both are stated, enabling alternative interpretations.


\paragraph{Exposure box results}

Because of the substantial differences found between our results and those in~\cite{guiseppe_2011} we decided to replicate their approach of using a radon source and small sealed exposure box. The main difference between this approach and the lab-based measurements is the very different surface to volume ratio and resulting equilibrium factors. Compared to the lab measurements only a limited number of exposures were performed.

Table~\ref{tab:all_length} summarizes the $d_{sum}$ results obtained when exposing Cu, SiPM, HDPE, and PTFE samples in a sealed, approximately $\rm 25\; cm\times 15\; cm\times 15\; cm$ stainless steel exposure box to high radon activities, derived from the Pylon radon source. The resulting surface to volume ratio of 55.2 m$^{-1}$ (including sample holder) is much bigger than for the lab measurements. This difference becomes important for surface attachment in case secular equilibrium is not assumed. The source was operated in flow-through mode. Care was taken to estimate the Rn activity contained in the box accounting for radon growth and turbulent gas mixing. We confirmed our radon activity model by dedicated measurements of the activity using the RAD7 detector. Data and calculated activities agree well.

As seen in table~\ref{tab:all_length}, the effective collection lengths observed this way are considerably shorter than those obtained in a larger clean room. This measurement approach would lead to an underestimate of the $^{210}$Pb surface activity after a given exposure of parts in some assembly lab. Our observations with the exposure box are in agreement with the results published in~\cite{guiseppe_2011}. For the example of copper, $1/d_{sum}$ is found to be proportional to the surface to volume ratio.


\subsection{Tests of the Jacobi Model}

The Jacobi model~\cite{jacobi_1972,postendoerfer_1978}, mentioned in several articles on the subject of radon progeny attachment to surfaces~\cite{leung_2005,stein_2018, morrison_2018}, allows to model chain disequilibrium for given environmental conditions. The tests performed in this study serve to understand whether model results are robust enough to be used to correct for different environmental conditions and to convert effective collections lengths into disequilibrium-corrected lengths.
We are not aware of previous tests of the accuracy of the model results. 

The model depends on multiple tuneable input parameters for which broad ranges can be found in the health physics literature. In this section we discuss our tuning of the model input parameters and compare the resulting model output with data. The model quantifies the disappearance of radon progeny atoms through several mechanisms: radioactive decay of the unattached (free) fraction in air, attachment to surfaces, attachment to aerosols (forming the attached component), removal by ventilation and filtration. 
The model, therefore, depends on branching fractions describing the relative importance of the different mechanisms. 
For our implementation we followed the detailed description given in reference~\cite{nazaroff_nero_1988}.
As a starting point, numerical values for the various branching ratios were taken from~\cite{nazaroff_nero_1988}.
However, the input parameters are typically only given within rather broad ranges. 

\begin{widetext}
\begin{table}[htb]
    \centering
    \begin{tabular}{l|l|l|l|l|l|l|l}
        Condition & 
        Jacobi model/ & 
        $d_{sum}$ [m] &
        $f_{Po} \cdot d_{Po}$ [m] &
        $f_{Po}$ &
        $f_{Pb}$ &
        $f_{Bi}$ &
        $f_{sum}$ [s] \\
         &
        Experiment &
         &
         &
         &
         &
         &
         \\
        \hline\hline
        Fan on high, & 
        Jacobi model    &
        $0.024-0.076$ &
        $0.024-0.074$ &
        $0.063-0.045$ &
        $(5.5-3.4) \cdot 10^{-4}$ &
        $(6.2-3.3) \cdot 10^{-6}$ &
        $18.3-12.9$ \\
        tent closed & 
        Experiment    &
        $0.074 \pm 0.008$ &
        $0.161 \pm 0.021$ &
        $0.010 \pm 0.002$ &
        $-$ &
        $-$ &
        $3.9 \pm 0.6$\\ \hline
        Fan on low,  &
        Jacobi model    &
        $0.096-0.164$ &
        $0.083-0.152$ &
        $0.226-0.094$ &
        $(1.1-0.25) \cdot 10^{-2}$ &
        $(7.2-0.8) \cdot 10^{-4}$ &
        $87.5-30.9$ \\
        tent closed & 
        Experiment    &
        $0.160 \pm 0.011$ &
        $0.189 \pm 0.021$ &
        $0.039 \pm 0.013$ &
        $-$ &
        $-$ &
        $44.5 \pm 20.5$\\ \hline
        Fan off, &
        Jacobi model    &
        $0.194-0.201$ &
        $0.128-0.178$ &
        $0.362-0.114$ &
        $0.077-0.007$ &
        $0.035-0.0022$ &
        $332.0-49.7$ \\
        tent closed & 
        Experiment    &
        $0.234 \pm 0.015$ &
        $0.206 \pm 0.018$ &
        $0.155 \pm 0.052$ &
        $-$ &
        $-$ &
        $490.1 \pm 210.8$\\ \hline
        Fan off, &
        Jacobi model    &
        $0.294-0.467$ &
        $0.172-0.354$ &
        $0.663-0.308$ &
        $0.462-0.107$ &
        $0.425-0.086$ &
        $1952.6-470.9$ \\
        tent open & 
        Experiment    &
        $0.438 \pm 0.034$ &
        $0.295 \pm 0.019$ &
        $0.201 \pm 0.082$ &
        $-$ &
        $-$ &
        $534.8 \pm 241.8$\\ \hline
        SNOLAB  &
        Jacobi model    &
        $0.257-0.699$ &
        $0.183-0.542$ &
        $0.491-0.331$ &
        $0.061-0.030$ &
        $0.009-0.003$ &
        $284.4-162.6$ \\
         & 
        Experiment    &
        $0.368 \pm 0.013$ &
        $-$ &
        $-$ &
        $-$ &
        $-$ &
        $-$\\ \hline
    \end{tabular}
    \begin{minipage}{7in}
    \caption{Comparison of the Jacobi model and experimental results obtained with copper in all exposure conditions. The range in values for Jacobi model represents different deposition velocities for $^{218}$Po (see text for details). The standard errors of the means are reported for $d_{sum}$ and $f_{Po} \cdot d_{Po}$, while the standard deviations are shown for $f_{Po}$ and $f_{sum}$. }
    \label{tab:Jacobi_model}
    \end{minipage}
\end{table}
\end{widetext}

For time periods when the SabreBPM2 radon progeny detector was available, we further compared calculated and measured equilibrium factors.
Using a blower with known air flow $\Phi$ (e.g. in $\rm m^3/s$), the SabreBPM2 detector collects radon progeny from the air onto a fine filter. It measures the $\alpha$ and $\beta$-decays of the collected unstable deposit, the summed contribution of the free and dust-attached fractions.

Using the same approach as discussed in section~\ref{sec:intro_attachment}, the instantaneous collection rate, or gain factor $g_i$, on the filter for nuclide $i$ is: $g_i=\Phi\cdot f_i\cdot \mathcal{A}_{Rn}\cdot \tau_i$. In steady state (at least 3 h operation for all nuclides, few minutes for $^{218}$Po), the constant $^{218}$Po and $^{214}$Bi activities on the filter are: $A_{Po}=\Phi\cdot \mathcal{A}_{Rn}\cdot f_{Po}\cdot \tau_{Po}$ and $A_{Bi}=\Phi\cdot \mathcal{A}_{Rn}\cdot \left( f_{Po}\cdot \tau_{Po} + f_{Pb}\cdot \tau_{Pb}+f_{Bi}\cdot \tau_{Bi} \right)$, respectively. Combined with the air flux and a measurement of $\mathcal{A}_{Rn}$ with the RAD7 detector, one can extract two equilibrium-related quantities from the appropriate activity ratios:
$f_{Po}\cdot \tau_{Po}$ and $f_{sum}=f_{Po}\cdot \tau_{Po}+f_{Pb}\cdot \tau_{Pb}+f_{Bi}\cdot \tau_{Bi}$.
These two quantities do not contain any collection lengths.
We decided to not rely on the measurement of $\beta$-decay rates, thus limiting the scope of the comparison. Moreover, the $^{212}$Bi (Th-chain) contribution to the $^{218}$Po counting rate was subtracted using the instrument-provided $^{212}$Po $\alpha$ rate.

Tuning of the Jacobi model parameters included varying the $^{218}$Po attachment speed ($v_{Po}=\frac{d_{Po}}{\tau_{Po}}$) between 5 and 22 $\frac{m}{h}$, as reported in~\cite{nazaroff_nero_1988}. Assuming diffusive transport (with all radon progeny having the same diffusion constant~\cite{schiller_1984}),  with the attachment dominated by $^{218}$Po, the $^{214}$Pb and $^{214}$Bi attachment speeds were assumed to scale like $\frac{1}{\sqrt{\tau}}$.
The modelling of attachment to aerosols was done by normalizing the particle size distribution found in reference~\cite{talbot_2016} to the particle count measured by us between 0.3 and 0.5 $\mu m$. Particle size dependent attachment coefficients were taken from reference~\cite{porstendorfer_1994}.

Table~\ref{tab:Jacobi_model} compares model-calculated quantities with measurement for copper, the material with the most measurements. This comparison is done for all exposure conditions and for the range of attachment speeds mentioned above. A calculation was also made for SNOLAB conditions, reported in~\cite{stein_2018} but using a corrected air exchange rate taken from~\cite{snolab_handbook_2006}.
The following measured and calculated quantities are compared: 1) the summed effective collection length $d_{sum}$; 2) the $^{218}$Po collection length $f_{Po}\cdot d_{Po}$; 3) the equilibrium factor for $^{218}$Po $f_{Po}$; and 4) the parameter $f_{sum}$. The latter two parameters could only be utilized for those conditions where SabreBPM2 data is available.

As can be seen from table~\ref{tab:Jacobi_model}, measured $d_{sum}$ values show reasonable agreement with the model at the high end of the attachment speed range. Using the same range, measured effective $^{218}$Po lengths agree in some conditions with the model-derived values. Also here the larger collection speeds tend to improve agreement between model and data.
Using the same tuning range, our measured $f_{Po}$ and $f_{sum}$ values do not reproduce those derived from the model well. 

Based on these observations, we consider the Jacobi model to provide qualitative guidance. The model reproduces general trends under the change of environmental conditions but quantitative agreement with the data, at least with our tuning, is rather limited. Because of this inability to model equilibrium factors accurately we are not attempting to correct our effective lengths with $f$-values 
 derived from the Jacobi model. 


\subsection{Disequilibrium-Corrected Collection Lenghts}
\begin{figure}[bbt]
\includegraphics[width=7.5cm]{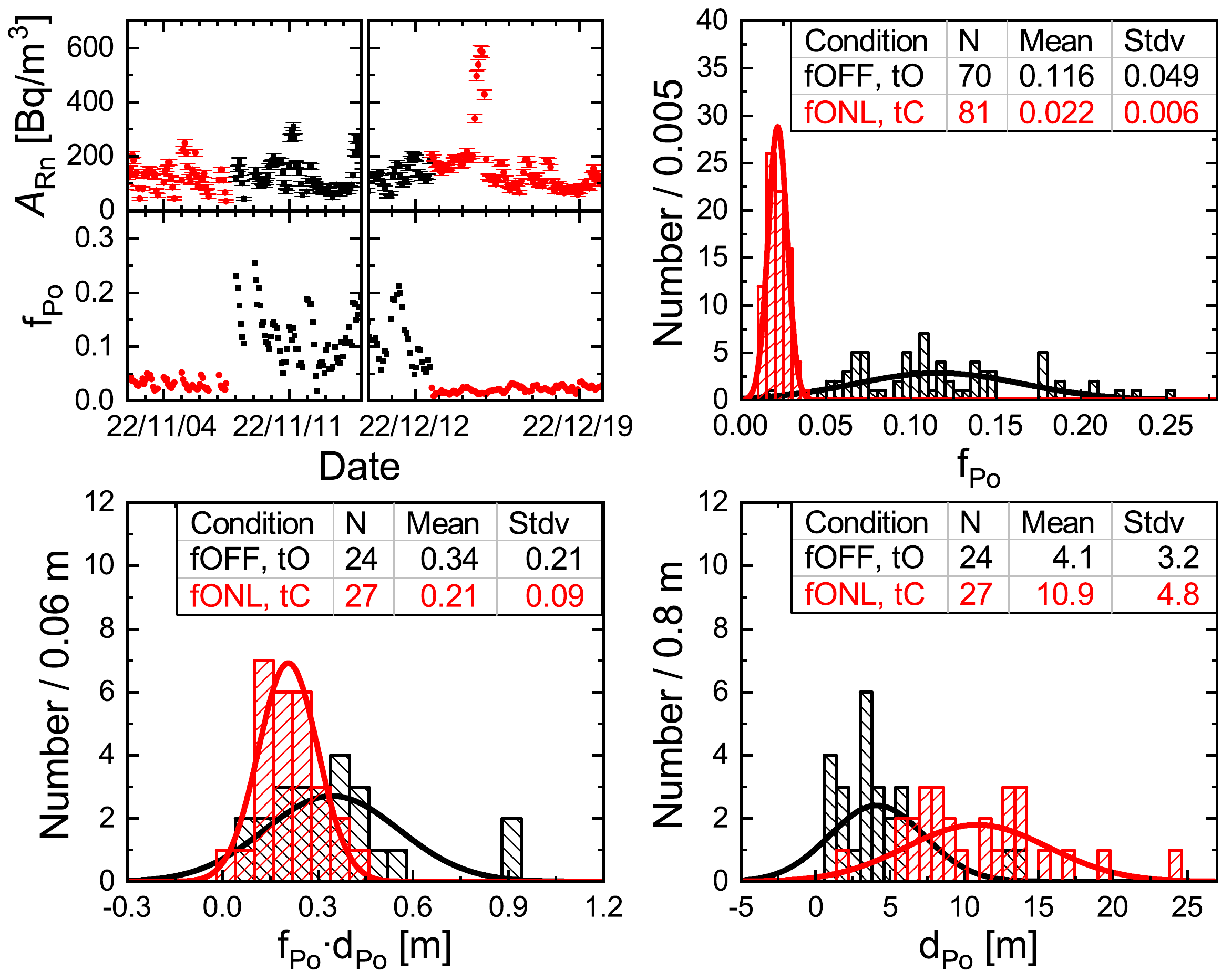}
\caption{\label{fig:dPo_lengths} 
Top left: Radon concentration and $^{218}$Po equilibrium factor $f_{Po}$ measured in the basement
lab using Durridge RAD7 and Bladewerx SabreBPM2 detectors. Environmental conditions were ``fan on low, tent closed'' before 2022/11/07 12:00, followed by ``fan off, tent open'' until 2022/12/12 13:30, and again ``fan on low, tent closed'' since then. Top right: Frequency distributions of $f_{Po}$ measured for the ``fan off, tent open'' (black) and ``fan on, tent closed'' (red) conditions. Bottom left: Frequency distribution of the $f_{Po} \cdot d_{Po}$ parameter obtained for 24 measurements with copper for ``fan off, tent open'' (black) and 27 copper measurements for ``fan on low, tent closed'' (red) conditions. Bottom right: Resulting frequency distributions of the disequilibrium-corrected deposition length $d_{Po}$ obtained by dividing each individual $f_{Po} \cdot d_{Po}$ parameter by the corresponding $f_{Po}$ value.
The same color coding as in the other panels is  used.}
\end{figure}
Using direct and simultaneous measurements of the $^{222}$Rn and $^{218}$Po volumetric activities in air, and of the $^{218}$Po surface activity on copper samples, we can determine $d_{Po}$ directly. The question of interest here is: does this length show environment dependencies beyond the differences in sub-chain equilibrium?

To obtain a direct measurement of $d_{Po}$, providing the dominant contribution to $d_{sum}$,
24 attachment measurements were performed in the basement lab in the “fan off, tent open” condition and 27 measurements in the “fan on low, tent closed” condition. The RAD7 radon detector and SabreBPM2 radon progeny monitor were operated simultaneously to sample over the same temporal variations of all environmental parameters. The top left panel of figure~\ref{fig:dPo_lengths} shows the $^{222}$Rn volumetric activities and measured $f_{Po}$ values during the two sampling periods. Only data obtained with an average radon concentration (of two consecutive measurements) above $50\; \rm Bq/m^3$ were included. The sharp increase in $f_{Po}$, seen in the bottom row of the top left panel (in black) of figure~\ref{fig:dPo_lengths}, coincides with the ventilation and filtration being turned off and the clean tent being opened. The impact air filtration, and with it chain equilibrium, has on the $^{218}$Po specific activity (being proportional to $f_{Po}$) is obvious. The upper right panel of figure~\ref{fig:dPo_lengths} shows how the presence of ventilation and HEPA filtration impact the equilibrium factor on average.

Comparing the two disequilibrium-corrected collections lengths, reported in the lower right panel of figure~\ref{fig:dPo_lengths}, shows that there is no unique length. Environmental differences impact radon progeny collection beyond the differences in chain equilibrium. The fact that the ordering of lengths is reversed by this correction may indicate the availability of a larger $^{218}$Po reservoir to surface attachment when using vigorous air circulation.

The step in $f_{Po}$ (top left panel, black symbols), coinciding with a change in ventilation, clearly shows that air filtration removes airborne $^{218}$Po and impacts sub-chain equilibrium. 
The measured average $f_{Po}$ value for “fan off, tent open” was $f_{Po}=0.116 \pm 0.049$, while for the “fan on low, tent closed” condition it was $0.022 \pm 0.006$ (standard deviations reported).
During the same time the volumetric $^{222}$Rn activity stayed more or less constant: $122 \pm 57 \frac{Bq}{m^3}$ and $154 \pm 109\; \frac{Bq}{m^3}$ (standard deviations reported), respectively. 

These measurements require the combination of data obtained by three different devices. Our results obtained with the RAD7 detector and the surface attachment counting are related directly or indirectly to measurement standards in our lab. The manufacturer-supplied calibration of the RAD7 detector was cross checked directly with our Pylon radon source. The counting of surface attached Po-radioactivity is linked by means of a Monte Carlo model to an activity calibrated $^{210}$Po source. For the volumetric $^{218}$Po activity, measured with the SabreBPM2 detector, no such cross check could be devised. This aspect of our data analysis, therefore, relies on the correctness of the manufacturer-supplied instrument calibration.

\subsection{What Factors Impact Collection?}
\begin{figure}[bt]
\includegraphics[width=7.5cm]{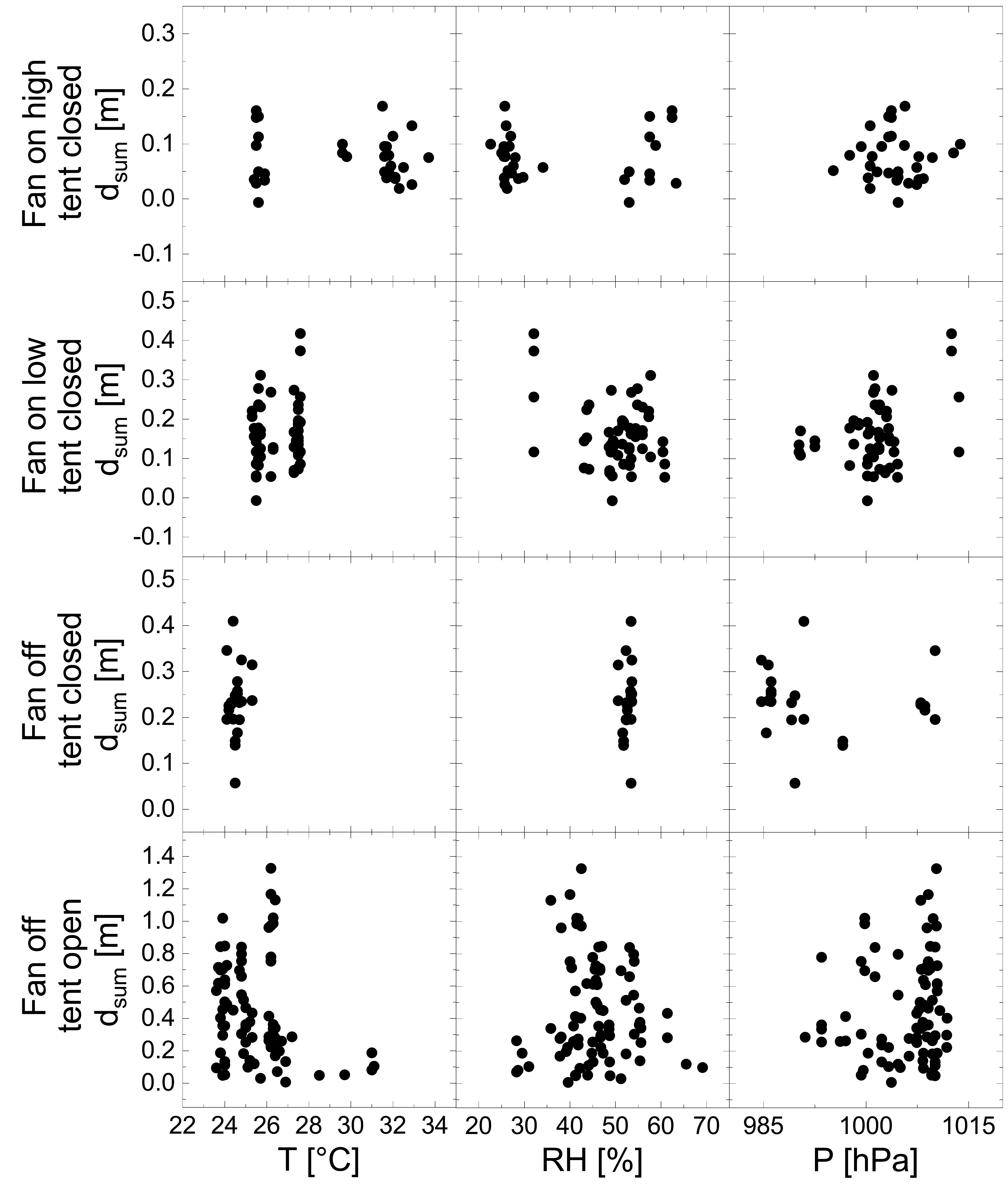}
\caption{\label{fig:cu_correlations} 
Example parameter correlation plots obtained for copper, the material with the most measurements. No clear correlations are observed. $T$ denotes temperature, $RH$ relative humidity and $P$ air pressure.}
\end{figure}
Environmental conditions in the collection area such as ambient temperature, pressure and humidity were monitored during the entire data taking period. The naturally occurring variation of these parameters was used to search for correlation with $d_{sum}$. The range of variation was 24-34$^{\circ}$C for temperature, 23-69\% for humidity, and 985-1014 hPa for air pressure.
Within these ranges no clear parameter correlation could be identified. As an example, figure~\ref{fig:cu_correlations} shows the correlation data for copper.

\begin{figure}[bt]
\includegraphics[width=7.5cm]{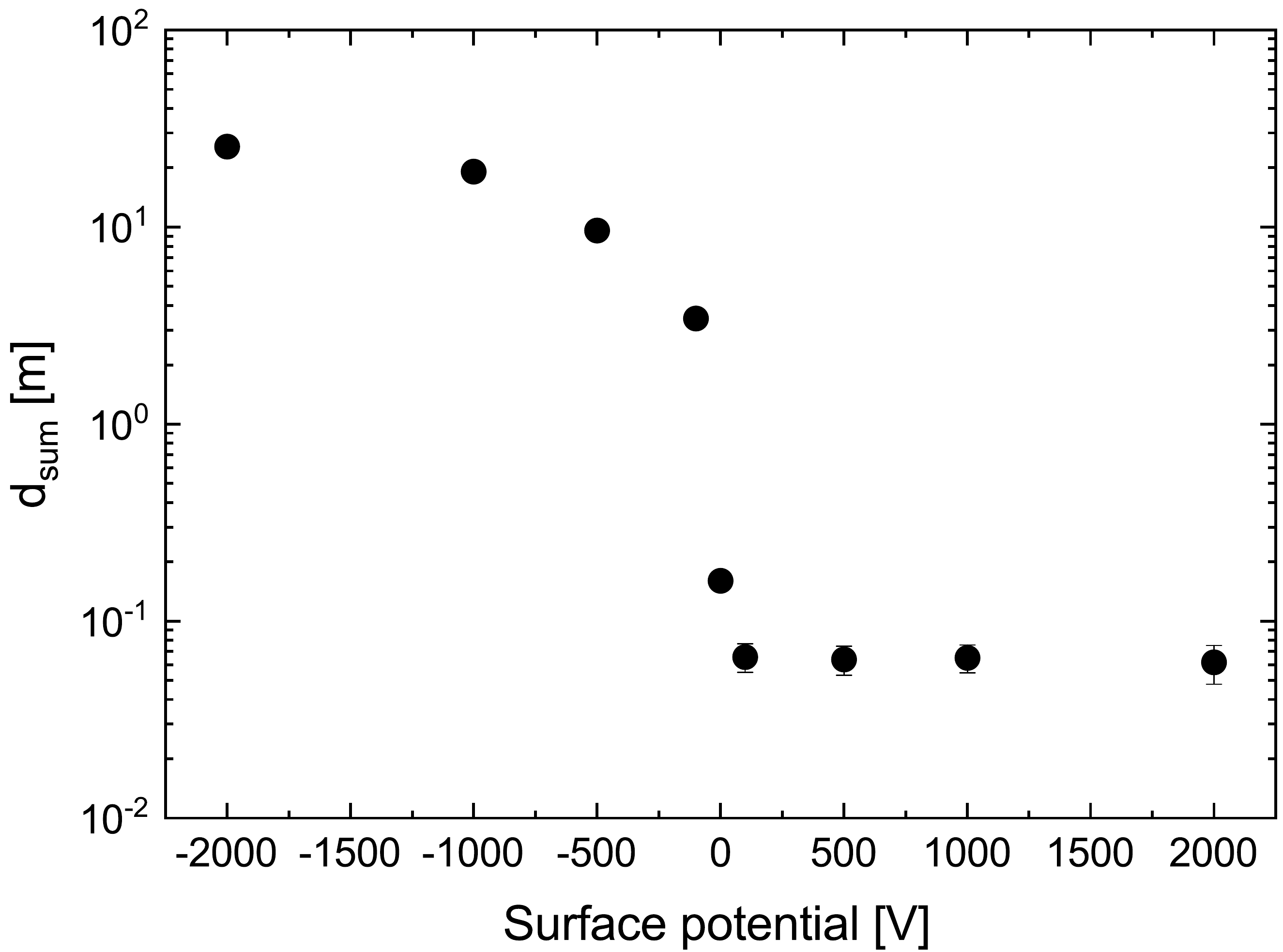}
\caption{\label{fig:cu_biased} 
Average effective collection lengths observed for Cu in the ``fan on low, tent closed'' condition. The sample was biased using a power supply. About 12 measurements were performed per voltage setting. The error bars correspond to the standard deviation of the data.}
\end{figure}
A parameter that does impact $d_{sum}$ is the presence of an electrical surface potential. This behaviour was already observed in~\cite{morrison_2018}. To study this effect we collected data with a copper sample, biased by a power supply to different voltages. About 12 exposures were performed for each of the 9 voltage settings. The voltage ranged from -2000 to 2000 V. The impact of biasing is clearly visible in figure~\ref{fig:cu_biased}. 
The saturation of the curve at positive biases indicates the presence of an electrically neutral population. The strong enhancement at negative bias identifies the dominant radon progeny charge state. It is not clear whether the saturation relates to radon progeny collection details or is due to the finite size of the collection room.

To test attachment suppression mechanisms, collection measurements were performed with a copper sample covered by a single Kimwipe tissue. Exposure tests with a covered copper sample showed no detectable activities, even in repeated exposures. To boost attachment, the tests with covered sample were repeated biasing the copper disks to -1000 V. The counting rates of $^{218}$Po and $^{214}$Po remained low, and only about 40$\%$ of acquired data was usable. The ratio of averaged $d_{sum}$-values of uncovered over covered biased copper samples was found to be $1059\pm 250$, surprisingly large given the simplicity and cost effectiveness of the suppression. Air tight wrapping was not needed, the Kimwipe served as an effective local sink for the radon decay products.

\section{Conclusion}
This paper presents a systematic study of radon progeny attachment to surfaces of multiple materials. The treatment of the problem does not use the problematic chain equilibrium assumption found in previous studies on this subject. The impact of environmental conditions on the attachment was investigated. We show that there is only little material dependence in the attachment lengths, that attachment is dominated by $^{218}$Po, that chain equilibrium and with it effective collection lengths depend strongly on the environmental conditions, namely air filtering, provide a test of Jacobi model calculations against data, and report the first equilibrium factor corrected measurement of the attachment length to copper.
The impact of surface biasing on radon progeny collection has been quantified.

\begin{center}
    {\bf Acknowledgements}
\end{center}
The work presented here was conceived as a contribution
to the development of the nEXO experiment. We would like
to thank our collaborators for stimulating and useful discussions.
This research was supported in part by the DOE Office of Nuclear Physics under grant number DE-FG02-01ER41166. We are thankful to Hamamatsu Photonics for providing SiPM samples. We are grateful to Dr.~R.~Tsang for valuable discussions and proofreading. We thank to Dr.~B.~Mong and C.~Kenney of SLAC for providing metallized silica samples, and A.~House of LLNL for carbon fiber composite samples. We are grateful to Dr.~V.~Zdimal, Dr.~N.~Talbot, Dr.~J.~Ondracek, Dr.~E.~Morrison, Dr.~R.~Schnee, Dr.~M.~Stein, and Dr.~D.~Jardin for providing data and valuable discussions. We thank D.~Baltz from Bladewerx LLC for help with the SabreBPM2 data analysis.

\bibliography{radon_progeny_attachment}

\end{document}